\documentclass[reprint,
superscriptaddress,
 amsmath,amssymb,
 aps,
prl,nofootinbib
]{revtex4-1}

\usepackage{graphicx}% Include figure 
\usepackage{dcolumn}% Align table columns on decimal point
\usepackage{bm}
\usepackage{hyperref}
\hypersetup{colorlinks = true, linkcolor = [rgb]{0.19411,0.51882,0.667058}, urlcolor = [rgb]{0.125490,0.29542,0.1647058}, citecolor = [rgb]{0.75882,0.37411,0.14117}}

\graphicspath{{../images/}}

\usepackage[toc,page]{appendix}

\newcommand{\nn}{\nonumber \\}
\newcommand{\bra}[1]{\langle{#1}|}
\newcommand{\ket}[1]{|{#1}\rangle}
\newcommand{\braket}[2]{\langle{#1}|{#2}\rangle}

\def\labell#1{\label{#1}}
\def\togli#1{}

\def\>{\rangle}
\def\<{\langle}

\usepackage{braket}

\def\>{\rangle}
\def\<{\langle}

\usepackage[bottom]{footmisc}

\begin{document}
% \fbox{{\scriptsize Preliminary draft \today}}
\title{Cryptographic quantum metrology}

\author{Zixin Huang}
\affiliation{ Department of Physics \& Astronomy, University of Sheffield, UK }
\affiliation{ School of Physics, University of Sydney, NSW 2006, Australia }

\author{Chiara Macchiavello}

\affiliation{Dip. Fisica and INFN Sez. Pavia, University of Pavia, via Bassi 6, I-27100 Pavia, Italy}

\author{Lorenzo Maccone}

\affiliation{Dip. Fisica and INFN Sez. Pavia, University of Pavia, via Bassi 6, I-27100 Pavia, Italy}

\begin{abstract}
  We develop a general framework for parameter estimation that allows
  only trusted parties to access the result and achieves optimal
  precision. 
  % The protocol is unconditionally secure if the trusted parties 
  % wish only to achieve the standard quantum limit. 
  % If Heisenberg-limited precision is to be achieved, 
  % they sacrifice some security -- 
  The protocols are designed such that adversaries can access some information 
  indeterministically, but only at the risk
  of getting caught (cheat-sensitivity); under the assumption that the adversary can access the channel only once,
  then the protocol is unconditionally secure. By combining techniques from quantum cryptography and quantum metrology, we devise
  cryptographic procedures for single parameter estimation when an
  arbitrary number of parties are involved.
   % and multiple parameter
  % estimation protocols for when one or two parties are involved. 
\end{abstract}
%\pacs{}
%quantum mechanics, 03.65.Ta
%quantum algorithms and protocols, 03.67.Ac
%quantum optics, 42.50.Gy
%Quantum noise, 42.50.Lc

\date{\today}

\maketitle

\section{Introduction}

Classical protocols for sharing measurement results, e.g.~secret
location sharing
\cite{cheng2014user,herrmann2014practical,dong2011longitude,puttaswamy2014preserving},
use classical encryption schemes that rely on assumptions such as a
bounded computational capacity of the adversaries. Quantum
cryptography \cite{bb84,scarani2009security,
 bennett1992quantum}
instead promises unconditional security: the only assumptions are the
laws of physics, and a correct implementation. Here we introduce a
general framework for quantum cryptographic protocols specifically
suited to the task of securing measurement outcomes (parameter
estimation) while retaining the highest available precision allowed by
quantum mechanics (quantum metrology). 
Clearly, one could perform
optimal parameter estimation and then use conventional quantum
cryptographic protocols to securely transmit the result. 
As we show
here, thanks to the quantum nature of the states employed in quantum
metrology, simple
modifications of conventional quantum metrology protocols allow secure
transmission of the estimated parameter.  
% This is particularly
 While a few such schemes have appeared in the literature
\cite{PhysRevA.65.022309,giovannetti2001quantum,giovannetti2002quantum,
  PhysRevLett.98.120501,PhysRevA.72.042338}, they were suited only to
specific cases. Here we give a general framework that can be adapted
to any quantum metrology protocol, which can be turned into a quantum-secured one with the prescriptions given below. For sub-shot-noise
estimation security here is
intended as cheat-sensitivity
\cite{PhysRevLett.92.157901,PhysRevLett.100.230502}: adversaries can
access information but only at the risk of being caught. This security
model is appropriate only for situations in which the penalty of being
caught is higher than the payoff of siphoning some information. The
presented protocols can also achieve unconditional security under the
hypothesis that Eve can interact with the probes only once.

Our goal is to securely and optimally estimate an arbitrary parameter
$\varphi$, encoded onto a probe through a unitary operator $U_\varphi
= e^{-i\varphi H}$, where $H$ is a known Hermitian operator, in an
ideal noiseless scenario.

In our framework, a trusted party, Charlie, holds the black box which
encodes the unitary $U_\varphi$.  He can switch 
between implementing $U_{\varphi+ \pi m/N}$ and $U_{\pi m/N}$,
where $m \in \{0,..,N-1 \}$
(we will see later on why this is needed). Charlie does not need to know
$\varphi$: he just has to add an additional phase in the first case and reroute
the probes in the second.
Charlie can classically
communicate with the other trusted parties (Alice, Bob), but cannot prepare 
quantum states.
 % -- otherwise Charlie can use conventional quantum
% cryptographic protocols to securely transmit the result. \textcolor{blue}{In reality,
Charlie can be a sensor at a remote location where the trusted parties
do not have easy access, e.g. a small device which collects data at a remote location, and has
limited experimental capabilities.

As is customary, we allow the eavesdropper
Eve complete control of the channel where the probes travel.  The main idea is simple:
in quantum metrology the measurement probes are prepared in an
entangled state (e.g.~the NOON state) which has the feature that
separate measurements on each probe give no information on the
parameter until they are jointly processed, because of the
entanglement. Moreover, a test of correlations on a complementary
observable of the probes can test for the presence of Eve as in
conventional quantum cryptography: any action by Eve will ruin the
correlation in at least one of two complementary properties. If she is
detected, the protocol is terminated. For example, in the secret estimation of distance between
two parties
\cite{PhysRevA.65.022309,giovannetti2001quantum,giovannetti2002quantum,PhysRevLett.98.120501,PhysRevA.72.042338}
$H$ and $\varphi$ represent the energy and the time of arrival of the
probes respectively, which are the two complementary observables that
must be tested to exclude the presence of Eve.

\begin{figure}[h]
\includegraphics[trim = 0cm 0cm 0cm 0cm, clip, width=0.66\linewidth]{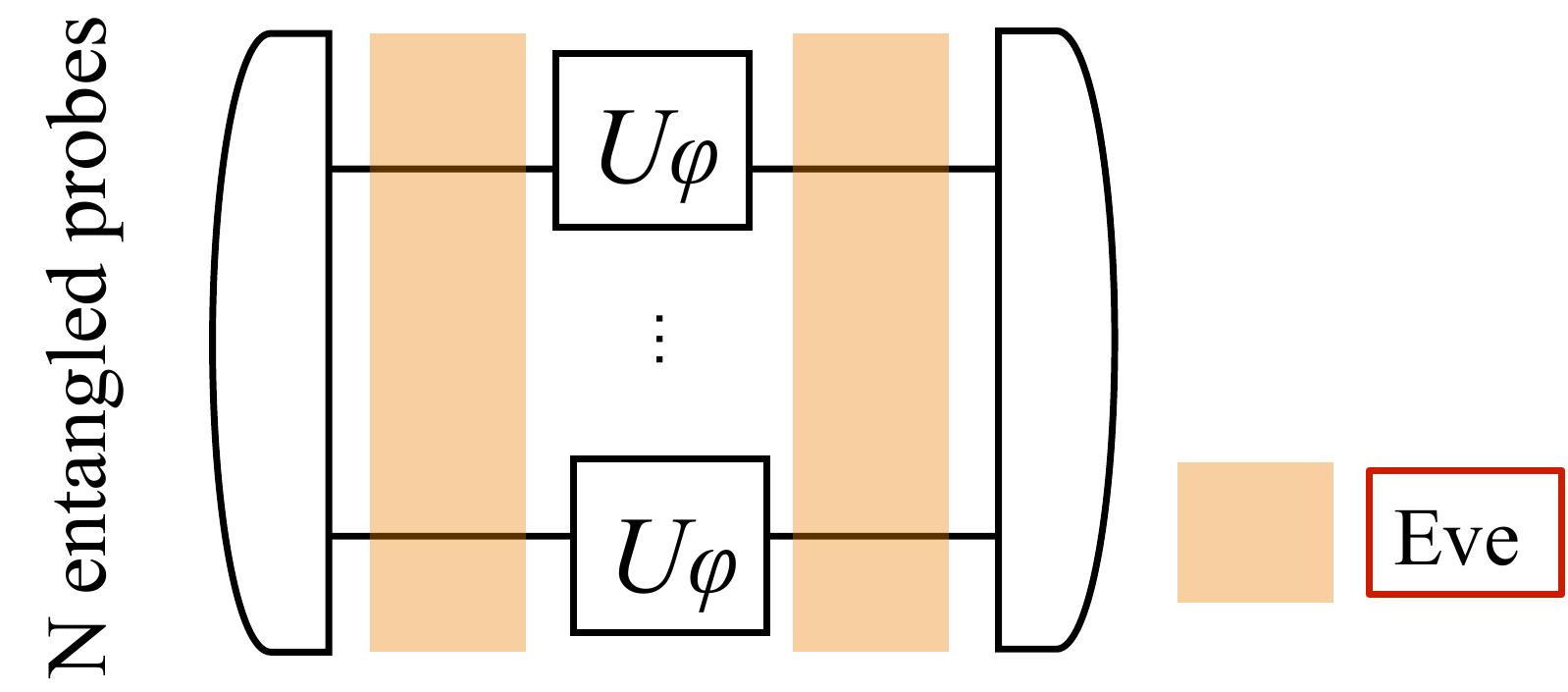} 
\vspace{-3mm}
  \caption{\label{f:scheme} A strategy which
    achieve the Heisenberg limit: the parallel-entangled strategy: a state of $N$ probes goes
    through $N$ maps in parallel. The channel $U_\varphi$ encodes the
    parameter to be estimated onto the probe states. The channels are
    also subjected to possible manipulation by an eavesdropper, Eve,
    denoted by the shaded regions.}
\end{figure}

The optimality of the parameter estimation is achieved through quantum
metrology
\cite{caves,giovannetti2004quantum,PhysRevLett.96.010401,PhysRevLett.98.090501}.
It establishes the best precision attainable in terms of the resources
devoted to it: if one is allowed $N$ uses of the transformation
$U_\varphi$, one can at most achieve the Heisenberg limit scaling of
$1/N^2$ in the variance (both in the finite dimensional
\cite{PhysRevLett.96.010401} and in the infinite dimensional
\cite{PhysRevLett.108.260405,PhysRevLett.108.210404} cases).  Among
the strategies to achieve the Heisenberg limit
\cite{PhysRevLett.96.010401,PhysRevLett.113.250801} here we use (i)
the sequential scheme where the map $U_\varphi$ acts on one probe $N$
times and (ii) the parallel entangled-scheme where an entangled state
of $N$ probes goes through $N$ maps $U_\varphi$ in parallel (see
Fig.~\ref{f:scheme}). In the latter case, entanglement among the $N$
probes is necessary, whereas separable states can only achieve the
standard quantum limit scaling of $1/N$ \cite{PhysRevLett.96.010401}.

The outline of the paper follows. In the first section, we summarize
the key results of quantum metrology, then detail how it can be turned
into cryptographic protocols involving one, two or an arbitrary number
of parties.  

\vspace{5mm}

Quantum metrology deals with the optimal estimation of a parameter
$\varphi$ which is encoded into a probe by a unitary map
$U_\varphi=\exp[-iH\varphi]$, where $H$ is the Hermitian generator.
The optimal initial states of the probe are the ones that have a
maximum spread for the generator: %for the sequential strategy where
% only one probe is employed its state is \mbox{$\ket{+} =
%   (\ket{\lambda_m} + \ket{\lambda_M} )/\sqrt{2} $ }, where
% $\ket{\lambda_m},\ket{\lambda_M}$ are the eigenvectors of $H$
% corresponding to the minimum and maximum eigenvalues; 
for the
parallel-entangled strategy an optimal state for the $N$ probes is the
NOON state \mbox{$ (\ket{\lambda_m}^{\otimes N} +
  \ket{\lambda_M}^{\otimes N} )/\sqrt{2}$}, where
$\ket{\lambda_m},\ket{\lambda_M}$ are the eigenvectors of $H$
corresponding to the minimum and maximum eigenvalues
\cite{Dowling2008,PhysRevLett.96.010401}. 
 After the evolution, the state %$N$ interactions
% the sequential strategy is transformed into
% $e^{iN\varphi% \lambda_m }|\lambda_m\>+e^{iN\varphi \lambda_M }|\lambda_M \>$ and is measured with the
% observable is %
% \mbox{$\hat O^\pm =(\ket{+}\bra{+} \pm \ket{-}\bra{-})$},
% where \mbox{$\ket{\pm} \propto|\lambda_m\>\pm|\lambda_M\>$}. 
% 
% state of the parallel strategy 
is transformed into
$e^{iN\varphi \lambda_m }\ket{\lambda_m}^{\otimes N} + e^{iN\varphi \lambda_M} \ket{\lambda_M}^{\otimes
  N} $ and can be measured ith the observable \mbox{$O_N^+ = (\hat
  O^+)^{\otimes N}$}, with $O^\pm=(\ket{\lambda_m}\pm\ket{\lambda_M})/\sqrt2$. The observable has an expectation value
$\cos[N\varphi(\lambda_M-\lambda_m)]$, whence one can estimate the
parameter $\varphi$. After repeating the estimation procedure $\nu$
times, the error in the estimation asymptotically in $\nu$ attains
the inequality
\cite{PhysRevLett.96.010401}
\begin{equation}
\Delta \varphi^2 \geq 1/\nu [N (\lambda_M- \lambda_m)]^2,
\label{eq:var}
\end{equation}
\noindent which corresponds to the Heisenberg limit. 
However, using purely N-probe NOON states only allows the phase to be estimated modulo $2\pi/N$
because there are $N$ fringes in $2\pi$.
 To resolve this ambiguity, smaller NOON states
with $N= 1,2,4...$ also have to be used, adding a small overhead
to the precision in Eq.~\eqref{eq:var}
\cite{PhysRevA.80.052114,higgins2009demonstrating,higgins2007entanglement}.
\color{black}

% \vskip 1\baselineskip
\section{Single party secure estimation \label{sec:1} }

Transforming a metrology protocol into a quantum cryptographically
secure one is simple in the one-party scenario, where a single party
(Alice) is in charge both of the preparation and measurement.

The protocol is designed such that Eve cannot extract information on
$\varphi$ or bias the measurement results without risking being caught (cheat sensitivity), even if she is in charge of
the channel between Alice and the maps $U_\varphi$. Namely, we assign
Eve the possibility of performing arbitrary joint transformations on
the probes both after Alice's preparation and before Alice's
measurement. If Eve biases the measurement by applying an additional phase $U_\theta,
 \theta \neq 0$, she needs to be discovered.

 The cryptographic protocol requires
Alice to choose randomly to prepare the phase-sensitive states $\ket{\Psi_N ^\pm}$,
 each with probability $P_a/2$,
and eigenstates of the Hamiltonian $\{ \ket{\lambda_0},\ket{\lambda_1} \}$
 each with probability $(1-P_a)/2$, where
\begin{eqnarray}
% \ket{\Psi_\text{probe}^\pm} &=& \ket{\Psi_N^\pm}^{\otimes 2}  \label{eq:probe}\\
\ket{\Psi_{N}^\pm}  &= &1/\sqrt{2} (\ket{\lambda_m}^{\otimes N}\pm \ket{\lambda_M}^{\otimes N})\\ 
 \ket{\lambda_0 }    &=&\ket{\lambda_m}^{\otimes N} \\
 \label{eq:lambdaa}
 \ket{\lambda_1 }    &=&\ket{\lambda_M}^{\otimes N}.
 \label{eq:lambdaab}
\end{eqnarray}

Alice sends through the probes one by one (there cannot be more than one probe in the channel at any one time), and only sends the next after the previous one returns\footnote{\color{black}For large $N$, if Alice sends the entire state through all at once, then Eve can easily estimate the phase unitary herself, apply her guess to Alice's state and send it back to Alice; here Eve will get away with much greater probability. }
Charlie keeps count of the number of probes going through the devices, and if there are too many many probes, it means that Eve is sending some of her own and they abort the protocol. On each entire state that Alice sends, he chooses to randomly implement $U_{\varphi+\pi m/N}$ and $U_{ m \pi/N}$ with probabilities $P_c$ and $(1-P_c)$ respectively.

Whenever Alice prepares $\ket{\lambda_{0/1}}$, or when Charlie applies $U_{ m\pi/N}$ and Alice had prepared $\ket{\Psi_N^\pm}$, Alice will end up with a decoy state. Here the term ``decoy" denotes a state that is not encoded with the parameter $\varphi$, which can be used to implement security checks.
 If she wants the complementary sets of decoy states to occur with equal probabilities, she chooses $P_a(2-P_c) =1$ (see supplementary material).

If Alice had prepared
$\ket{\Psi_{N}^+}$ or $\ket{\Psi_{N}^-}$, after
the $U_\varphi$ interactions, she measures the observable $\hat O_{N}^+$
or $\hat O_{N} ^- $ respectively, separately on each of the $\ket{\Psi_{N}^\pm}$. 
Instead, when she had prepared $\ket{\lambda_0},\ket{\lambda_1}$, she performs a projective measurement back onto
this basis. 
% \color{blue}
If she had prepared $\ket{\Psi_N^\pm}$ and finds that the measurement
does not match the preparation, she informs Charlie, which will in turn reveal
whether he applied the check $U_{m\pi/N}$. Projecting a probe state back onto the NOON basis remains deterministic, since
\begin{eqnarray}
U_{m\pi/N} \ket{\Psi_{N}^\pm} = 
  \begin{cases}
    \ket{\Psi_{N}^\pm}, & \text{if } m \text{ is even,}\\
    \ket{\Psi_{N}^\mp}, & \text{if } m \text{ is odd}.
\end{cases}
\label{eq:cases}
\end{eqnarray}
From the states which Charlie 
has applied the check unitary, they can deduce whether Eve has
biased the measurement.

% \color{blue}
Now, we design the protocol such that the channel is a dephasing one if $m$ is unknown. 
This can be achieved by Charlie implementing $U_{\varphi+ m\pi/N}$ with random $m \in [0,..,N-1]$.
 If we limit $m$ (say $m=0,1$), then Eve could 
 be using single qubits instead of a N00N state to get an estimate. 
 For large $N$, $\pi/N$ becomes small. If Eve replaces Alice's probes with her own
and she knows the value of N, then if she sees anything other than the identity or $\pi/N$, 
she will know Charlie may have applied the parameter and gain a decent estimate on $\varphi$.

\color{black}
For both the check cases, the outcomes are deterministic.
If the measurement outcome does not match the state preparation/evolution,
Alice knows Eve has been tampering and the protocol is terminated, or 
they can estimate Eve's bias. 

As the last step in the protocol,
if the check shows that the process has been noiseless, only then
Charlie reveals on which probes he applied which unitary, i.e. the value
of $m$ in $U_{\varphi+\pi m/N}$.
Alice then
computes the observable $\braket{ \hat O_{N}^+} + \braket{\hat O_{N} ^-}=
\cos[(N\varphi + m\pi) (\lambda_M- \lambda_m)]$, whence she obtains $\varphi$.
All the public communication is useless to a third party.

Asymptotically, the achievable root-mean-square error is 
% \color{blue}
\mbox{$\Delta\varphi^2 \geq 1/P_a P_c \nu [N(\lambda_M-\lambda_m)]^2$}, \color{black}
 because with probability $P_c$ Charlie applies $U_{m \pi /N}$.
%   they achieve quantum Fisher information $N^2/2$. Then, on average 
% $1-\eta$ fraction of the states are employed as decoys:

% 
% If large NOON states are not available, Alice can evolve the $N=1$ state $N$ times sequentially in the channel and achieve the same precision (Fig.~\ref{f:scheme} (i)). This is at the expense of her waiting time being increased by a factor $N$.

% \color{blue}

With probability $P=1-P_a P_c$, Alice would have a decoy 
state at hand, and if Eve tampers with the estimation, she will be discovered with
probability $(1-\frac{P}{4})^\kappa$, where $\kappa$ is the number of states she tampers with.

% Even if Eve has complete control of the channel, the protocol is
% unconditionally secure (that is, she gains information on $\varphi$ with
% exponentially small probability). For example, for $N=1$, with $m = 0$ and 1 occurring with equal probability,
% the channel 
% % 
% \begin{eqnarray}
% \rho' \rightarrow \frac{1-P_c}{2}(\rho + \sigma_z \rho \sigma_z^\dagger) + \frac{P_c}{2}(U_\varphi \rho U_\varphi ^\dagger +  U_{\varphi+\pi} \rho  U_{\varphi+\pi}^\dagger) \nn
% \end{eqnarray}
% \noindent is equivalent to the completely dephasing channel.
% For $N=2$, the protocol is also unconditionally secure. If Eve chooses to use $\ket{\Psi_2^\pm}$, then the problem maps exactly onto the $N=1$ case. If she chooses to use $\ket{\pm}$ twice, the possible outcomes are $++, +-, -+, --$, which have no resolution in distinguishing between $I, e^{i \pi/2}$ and $\varphi$, since applying $e^{i \pi/2}$ can result in all the possible measurements.

% \color{blue}
Eve cannot gain information on $\varphi$ deterministically - to estimate $\varphi$, Eve needs to have hijacked Alice's probes whilst
the phase unitary is applied, and not having been caught previously. If she is discovered,
Charlie keeps the value of $m$ to himself, and Eve estimates $\varphi + \pi m/N$ with an unknown $m$, 
which is useless.

The maximum QFI Eve gains is $\kappa N^2$, where $\kappa$ is the number of probe states she tampers with, and this occurs with exponentially small probability
$(P_c) ^\kappa $.

%  Eve's information gain will then scale as a
% negative exponential with $\nu$, although it scales linearly with $N$, since
% she just has to successfully cheat a few times to achieve linearity in
% $N$. 

 Note that, if Eve cannot access the channel twice (and can only
 attempt to recover $\varphi$ by measuring Alice's probes), then the
protocol is unconditionally secure, since from her point of view the
state propagating in the channel is a mixed state,
\begin{align}
  [(|\lambda_m\>\<\lambda_m|)^{\otimes
    N}+(|\lambda_M\>\<\lambda_M|)^{\otimes N}]/2
\labell{eve}\;,
\end{align}
which is useless for parameter estimation since it does not acquire
any phase during the interaction $U_\varphi$.

\section{Multiple-party estimation}

In two-party protocols Alice is in charge of the state preparation and
Bob is responsible for the measurements
 (e.g.~a distance measurement
using light pulses and synchronized clocks). 
 They both wish to recover
the parameter in a way that is at least cheat-sensitive.
The procedure is inspired by the BB84 protocol.

The state preparation is the same as
for the single party protocol: Alice chooses randomly to prepare  
$\ket{\Psi^\pm _N}$ each with probability $P_a/2$, and
$ \ket{\lambda_{0/1}}$ each with
 probability $(1-P_a)/2$. 
% 
% 
% 
%  
% \color{blue}
 Bob independently chooses to measure either $\hat
O_{N}^+$ with probability $(1-P_a)/(1-P_a P_c)$, or a projective measurement onto $\{\ket{\lambda_{0}},\ket{\lambda_{1}}\}$, with probability $1- (1-P_a)/(1-P_a P_c)$ (this makes the probability of the decoy states symmetric, see supplementary material). 

\begin{figure}[hbt]
\includegraphics[trim = 0.0cm 0cm 0cm 0cm , clip, width=0.9\linewidth]{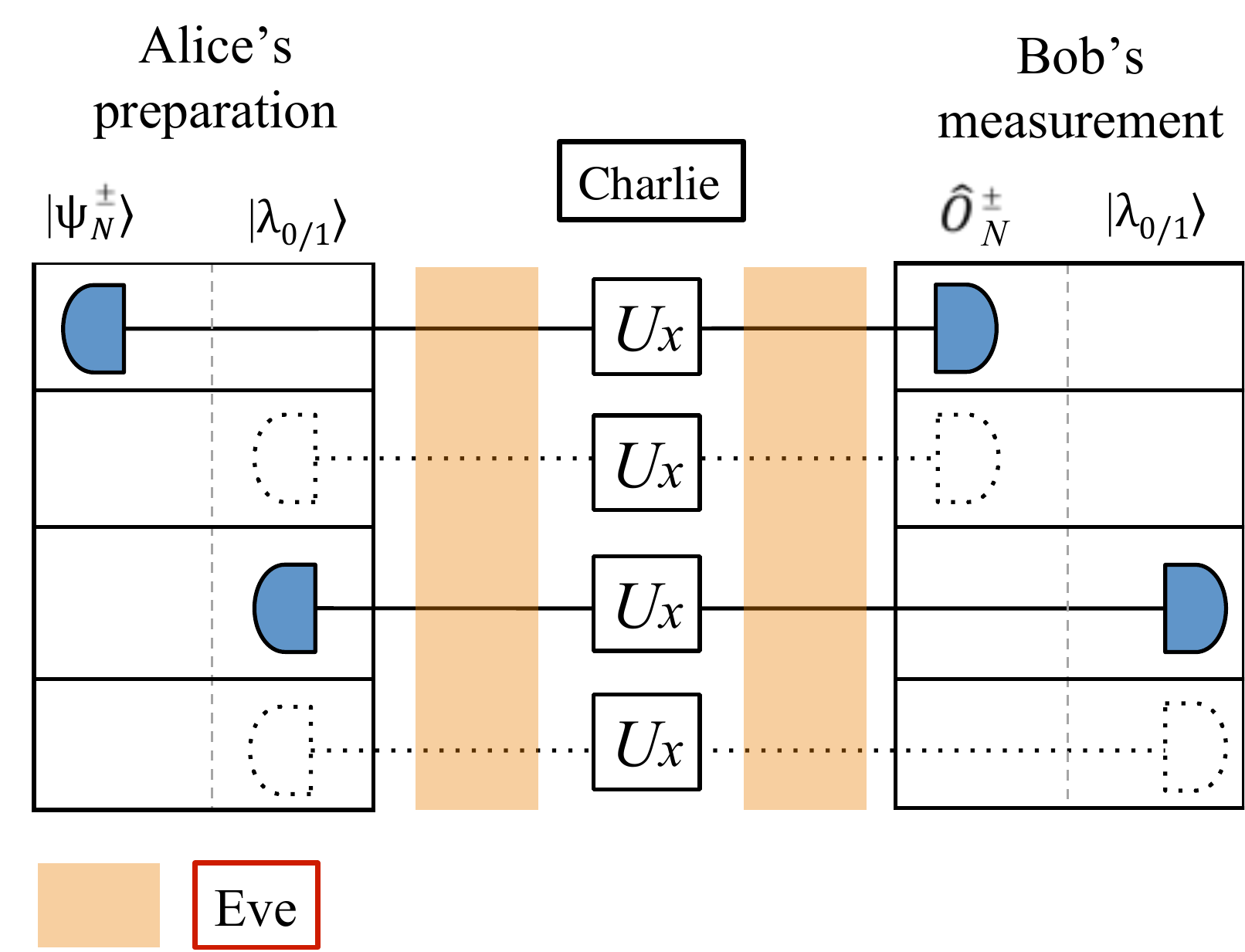}
% \vspace{-3mm}
\caption{\label{f:scheme1} Alice sends either 
$\ket{\Psi_N^\pm}$ or $\ket{\lambda_{0/1}}$ into the quantum channel which encodes the parameter $\varphi$ onto the probes. Charlie implements the unitary
 \mbox{$U_x, x \in \{ \varphi+ \frac{\pi m}{N}, m\pi/N \}$}. Bob randomly chooses to measure the observable $\hat O_{N} ^\pm$ or projects onto one of the basis of the decoy states. They retain only the copies for which their choice of basis agree, denoted by the solid blue markers. The probes are also subjected to possible manipulation by an eavesdropper, Eve, denoted by regions shaded orange.}
% \vspace{-3mm}
\end{figure}

Whenever Alice has prepared $\ket{\lambda_{0/1}}$, Charlie has applied the check unitary and Bob has measured in the correct basis,
namely, when the states Bob received are $\ket{\Psi_N^\pm}, \ket{\lambda_{0/1}}$,
the outcomes are deterministic.
\color{black}
After Bob's measurement, they use a public channel to
check their choice of measurement basis and discard all the cases when
they do not agree (see Fig.~\ref{f:scheme1}).

The exchange of classical information can be done at the end of the protocol, follows:
\begin{enumerate}
\item Alice reveals the basis of state preparation, Alice and Bob check for correlations on $\ket{\lambda_{0/1}}$. If the correlations are perfect, they proceed.
\item Charlie reveals on which states he has applied the check unitary $U_{ m \pi/N}$.
\item On the states Charlie has applied the check unitaries, Alice reveals which of the $\ket{\Psi_N ^\pm}$ she prepared, and they check correlations in this basis as in Eq.~\eqref{eq:cases}. If they decided that no one has tampered with the communication, Charlie discloses the values of $m$ on the states which he applied $U_{\varphi + \frac{\pi m}{N}}$.
\item Alice reveals on half the probe states whether she prepared $\ket{\Psi_N^+}$ or $\ket{\Psi_N^-}$, and Bob reveals his measurement outcomes on the other half.
% discloses on which copies he has applied which phase unitary. 
% 
\end{enumerate}

% Once they confirm that their communication is secure, Alice announces
% whether she has prepared $\ket{\Psi_N^+}$ or
% $\ket{\Psi_N^-}$ on half the copies of the probe states,
% and Bob reveals his measurement outcomes on the other half, and Charlie 
% discloses on which copies he has applied which phase unitary. 

They now can  compute $\hat O _{N} ^\pm$ correspondingly, where the sum of the
expectation values is once again \mbox{$\cos[(N\varphi + m\pi)(\lambda_M -\lambda_m)]$}, whence they can both obtain $\varphi$.
Four states are necessary for the two-party protocol, because
alternating between the plus or minus probe state ensures that their
communication is meaningless to a third party.

 The probability of Eve being undetected is (see Supplementary Material)
\begin{eqnarray}
\left(1- \frac{2 (1-P_a) P_a (1-P_c)}{4(1-P_a P_c)} \right)^\kappa.
\label{eq:twopartyprobs}
\end{eqnarray} 
This can be improved if Alice and Bob share a secret bit string in advance such that Bob knows which basis to choose, in which case Eve's probability of being undetected is  $(1- \frac{1-P_a P_c}{4})^\kappa$.
In addition, if Eve only has access to one end of
the channel, then she cannot gain any information, even if she
intercepts all the probes and the communication between Alice and Bob.

The achievable precision on $\varphi$ for each party
is
% The two-party protocol achieves a quantum Fisher information yield of $y  = \frac{1-P_a}{1-P_a P_c} P_a$ with respect to the optimal
% estimation, but the achievable precision on  %\mbox{$\Delta \varphi^2 \geq 16/\nu  N^2 $}
% \color{blue}
 \mbox{$\Delta \varphi^2 \geq 2/(\nu  \frac{1-P_a}{1-P_a P_c} P_a P_c [N(\lambda_M-\lambda_m)]^2 )$}.  
\color{black}
This is due to the
fact that, $P_a$ fraction of the time Alice sends out a phase-sensitive state,
 $\frac{1-P_a}{1-P_a P_c}$ fraction of the time Bob measures in the correct basis,
$P_c$ fraction of the time Charlie applies the phase unitary, and
 and each party estimates the parameter from only half the
remaining copies of the probe states. This reduction is only
a constant factor. For small $N$ the efficiency of the scheme can be improved by
using techniques such as those described in
Ref.~\cite{lo2005efficient}.

% \color{blue}{
The difference between the two-party protocol described above, and one where Alice simply performs the estimation and encrypts/sends it via quantum key is that this protocol can be tailored to the scenario where Alice (or Bob) does not learn the parameter. In the former protocol, or in any classical protocol this is impossible.

\color{black}
\vskip 1\baselineskip We now examine the multiple party scenario.
Alice and Charlie wish to measure and transmit the parameter to some trusted
parties, but she wants them to uncover the parameter only when they meet and collaborate, analogously to quantum-secret-sharing schemes
\cite{PhysRevA.59.1829,PhysRevA.61.042311,PhysRevA.71.044301,karlsson1999quantum,xiao2004efficient,cleve1999share}.
Here Alice is in charge of state preparation and we assume that
$\varphi$ is encoded by Charlie in the channel that
separates her from Bob.  If the secret is to be shared among $k$
trusted parties excluding Alice, she prepares
$\ket{\Phi_N ^\pm}$ with probability $P_a$ and $\ket{\Lambda_0},\ket{\Lambda_1}$ each with probability $(1-P_a)/2$, where
% \vspace{-5mm}
\begin{eqnarray}
 \ket{\Phi_{N}^\pm} = \frac{1}{\sqrt{2}}(\ket{\lambda_m}^{\otimes N+k-1} \pm 
  \ket{\lambda_M}^{\otimes N+k-1})  
\end{eqnarray}
and 
\begin{eqnarray}
\ket{\Lambda_0} = \ket{\lambda_m}^{\otimes N+k-1}, \qquad
\ket{\Lambda_1} =    \ket{\lambda_M}^{\otimes N+k-1}.
\end{eqnarray}
\color{black}
When she prepares \mbox{$\ket{\Phi_N^\pm}$}, she sends $N$ probes
from the state into the quantum channel to Bob,
 and one each to the other $k-1$ parties. The state $\ket{\Phi_N^\pm}$ evolves to
\begin{eqnarray}
\ket{\Phi_N^\pm} \rightarrow \frac{1}{\sqrt{2}}( &&
  e^{i (N\varphi + m\pi)\lambda_m}
  \ket{\lambda_m }^{\otimes N+k-1} \nn
&& \pm e^{i (N\varphi + m\pi) \lambda_M}
 \ket{\lambda_M}^{\otimes N+k-1}).
\end{eqnarray} 
  Now, if every party independently chooses randomly an observable to measure with
probability $\eta$, the scheme would be exponentially inefficient.
 % as the probability of them all matching is $\eta(\eta+\frac{1-\eta}{2})^{k-1}$
To overcome this, they
need to first agree on a sequence of measurement basis in a secure
way: Alice can perform a BB84 quantum key distribution separately with
each participant. Then she will share a unique secret bit string with
each of them. She then compares these bit strings, uses one as a
reference and instructs the rest to match theirs to it by performing a
series of bit flip operations.  This is a secure step as she is just
instructing which bit to flip, never communicating the bit's initial
or final value.  
Alternatively, multipartite conference key distribution protocol can 
be employed in order to establish a shared secret key \cite{epping2017multi}.

The parties then agree to project onto the $\ket{\pm}$
bases
at the $j$th iteration of the
protocol if the $j$th two-bit value is 0, given Alice will
send $\ket{\Phi_N}$.
If the bit value is 1, then they project onto the computational basis , as Alice will send $\ket{\Lambda_{0/1}}$.
Measuring $\hat O^\pm_{N+k-1}$ on $\ket{\Phi_N^\pm}$ will yield the outcome $\pm 1$.

\color{black}
As Alice sends through the states, they check the outcomes on the
decoys: if the measurements of all $k$ parties do not match Alice's
preparation, they know an eavesdropper is present and they abort the
protocol.

The rest of the protocol then follows trivially from the two-party version.
At the end of the protocol (this stage can be delayed arbitrarily),
Alice announces whether she prepared $\ket{\Phi_N ^+}$ or
$\ket{\Phi_N^-}$ on half the copies, and the rest of the
parties reveals their respective measurement outcomes on the other
half. Alice now possesses information on all the probes, and can
deduce the parameter by computing the observable $\hat O_{N+k-1}
=(\ket{+}\bra{+} - \ket{-}\bra{-})^{\otimes N+k-1}$, which has
expectation value
$\cos[(N \varphi + m\pi)(\lambda_M-\lambda_m)]$.  The precision of her estimate
is $2/\nu P_a P_c  [N(\lambda_M-\lambda_N)]^2$. \color{black}
  As for the other parties, they
now need to correlate their measurement outcomes in order to uncover
$\varphi$.  They do so by also calculating $\braket{\hat O_{N/2+k-1}} $.
Without information from any of the participants, the rest of the
results are useless, since this would be equivalent to tracing out one
probe from a maximally entangled state, which renders the measurement
outcome of the rest completely random.
The additional resources used are of $2k\nu$
qubits used for quantum key distribution and the $\nu(k-1)$ extra probes
that do not interact with $U_\varphi$.

\section{Conclusion}
By combining techniques from quantum metrology and quantum cryptography, we have defined a general framework for quantum cryptographic protocols specifically
suited to the task of securing parameter
estimation while retaining the highest available precision. Adversaries can gain some information on the parameter, but at the risk of being detected. We devised protocols for single parameter estimation involving an arbitrary number of parties.

\section{Acknowledgement}
We acknowledge funding from Unipv ``Blue sky'' project-grant
n.~BSR1718573.

\appendix
\renewcommand\thefigure{\thesection.\arabic{figure}}    
\setcounter{figure}{0}

\clearpage

\section{Supplementary information}

% \subsection{}

For single parameter estimation, Alice prepares the
probe state $\ket{\Psi_N^\pm}$ with probability $P_a$,
and the decoy states $\{ \ket{\lambda_{0/1}} \}$ each
with equal probability $\frac{1-P_a}{2}$. 

With probability $P_c$ Charlie applies the phase unitary
$U_{\varphi+m\pi/N}, m\in \{0,..., N-1\}$. The probability of the respective
states that Alice has are:

\begin{eqnarray}
P_a P_c &:& 1/\sqrt2 (e^{i (N\varphi + m\pi) \lambda_m}\ket{\lambda_m}^{\otimes N} \nn
        & & \pm e^{i(N\varphi + m\pi) \lambda_M} \ket{\lambda_M}^{\otimes N}) \nn
P_a (1-P_c)&:&1/\sqrt2 (\ket{\lambda_m}^{\otimes N}\pm  \ket{\lambda_M}^{\otimes N}) \nn
(1-P_a)&:& \ket{\lambda_0},\ket{\lambda_1} 
\label{eq:probs}
\end{eqnarray}

To maximise security, the decoy states $1/\sqrt2 (\ket{\lambda_m}^{\otimes N}\pm  \ket{\lambda_M}^{\otimes N})$ and $\ket{\lambda_{0/1}}$ need to occur with equal probability. This means that 
\begin{eqnarray}
P_a (1-P_c) = (1-P_a), \rightarrow P_a(2-P_c)=1.
\end{eqnarray}

If Eve wants to minimize her probability of getting caught while still obtaining some information, her best strategy would be to try to discriminate which decoy state Alice has prepared, and send this to Bob. For the decoy states we have considered,  the maximum discrimination probability is $2/S$, where $S$ is the number of symmetric states in the set \cite{PhysRevA.77.044302}. 

Eve can achieve this as followed: since she does not get caught if Alice prepares a probe state, Eve always makes a guess on the decoy state basis that Alice chose and performs a von Neumann measurement in that basis. When a decoy state is prepared, Eve will make an incorrect guess $1/2$ of the time. When she is correct, the state she sends back to Alice will correlate with Alice's preparation and Eve is undetected. When Eve is incorrect, she sends a state in the wrong basis -- but half the time, when Alice performs a measurement, it will collapse to be what she prepared. Therefore in a one-party scenario, the probability of Eve's cheat undetected is $(1- \frac{(1-P_a P_c)}{4})^\kappa$.

To increase the probability of detecting Eve whilst keeping the same efficiency $\eta$, Alice can use more complementary bases sets in the decoy state space, for example, including also 
\begin{eqnarray}
1/\sqrt{2} \left(\ket{\lambda_m}^{\otimes N} \pm i \ket{\lambda_M}^{\otimes N}  \right)
\end{eqnarray}
\noindent in the basis set. In this example, the probability that Eve will incorrectly guess the basis is now 2/3.

In the two-party protocol, the probability that Bob receives the states are the same as those in Eq.~\eqref{eq:probs}.
If Bob chooses the bases $\ket{\pm}$ and $\ket{\lambda_{0/1}}$ with probability $\eta,1-\eta$ respectively, then for the decoy states to occur with equal probability,
\begin{eqnarray} 
P_a (1-P_c)\eta & =&( 1-P_a) (1-\eta) \\
\rightarrow \eta &=&\frac{1-P_a}{1-P_a P_c}
\end{eqnarray}

Now, the probability that Bob has received a decoy state and measured in the correct basis is given by
\begin{eqnarray} 
P_a (1-P_c)\eta + ( 1-P_a) (1-\eta) = 
 \frac{2 (1-P_a) P_a (1-P_c)}{(1-P_a P_c)} . \nn
\end{eqnarray}

\noindent The probability of Eve successfully evading detection would be $1/4$, when this occurs, therefore if Eve tampers with $\kappa$ probes, the probability is given by 
\begin{eqnarray}
\left( 1-\frac{2 (1-P_a) P_a (1-P_c)}{4(1-P_a P_c)} \right)^\kappa.
\end{eqnarray}

\end{document}